\documentclass[
reprint,
superscriptaddress,
amsmath,
amssymb,
aps,
prl
]{revtex4-1}

\usepackage{graphicx}
\usepackage{xcolor} 
\newcommand{\ket}{\rangle}

\begin{document}

\title{Observation of Collisions between Two Ultracold Ground-State CaF Molecules}

\author{Lawrence W. Cheuk}
\email{lcheuk@princeton.edu} 
\affiliation{Department of Physics, Harvard University, Cambridge, MA 02138, USA}
\affiliation{Harvard-MIT Center for Ultracold Atoms, Cambridge, MA 02138, USA}
\affiliation{Department of Physics, Princeton University, Princeton, NJ 08544, USA}
\author{Lo\"ic Anderegg}
\author{Yicheng Bao}
\author{Sean Burchesky}
\author{Scarlett Yu}
\affiliation{Department of Physics, Harvard University, Cambridge, MA 02138, USA}
\affiliation{Harvard-MIT Center for Ultracold Atoms, Cambridge, MA 02138, USA}

\author{Wolfgang Ketterle}
\affiliation{Harvard-MIT Center for Ultracold Atoms, Cambridge, MA 02138, USA}
\affiliation{Department of Physics, Massachusetts Institute of Technology, Cambridge, MA 02139, USA }

\author{Kang-Kuen Ni} 
\affiliation{Department of Physics, Harvard University, Cambridge, MA 02138, USA}
\affiliation{Harvard-MIT Center for Ultracold Atoms, Cambridge, MA 02138, USA}
\affiliation{Department of Chemistry and Chemical Biology, Harvard University, Cambridge, MA 02138, USA}

\author{John M. Doyle} 
\affiliation{Department of Physics, Harvard University, Cambridge, MA 02138, USA}
\affiliation{Harvard-MIT Center for Ultracold Atoms, Cambridge, MA 02138, USA}

\date{\today}
\begin{abstract}
We measure inelastic collisions between ultracold CaF molecules by combining two optical tweezers, each containing a single molecule. We observe collisions between $^2\Sigma$ CaF molecules in the absolute ground state $|X,v=0, N=0,F=0\rangle$, and in excited hyperfine and rotational states. In the absolute ground state, we find a two-body loss rate of {$7(4) \times 10^{-11} \text{cm}^{3}/\text{s}$}, which is below, but close to the predicted universal loss rate.

\end{abstract}

\maketitle
The rich internal structure of molecules has led to many proposed applications ranging from precision measurements that probe beyond Standard Model physics to quantum simulation of condensed matter models and quantum computation~\cite{Carr2009review}. In the past decade, experimental advances in the production and control of molecules have brought many of the proposed applications within reach. These advances include coherent assembly of diatomic molecules from ultracold atoms~\cite{Ni2008KRb,Takekoshi2014RbCs,Park2015NaK,Gregory2016RbCs,Guo2017NaRb,Rvachov2017LiNa}, laser-cooling of molecules to $\mu$K temperatures~\cite{barry14,truppe17,anderegg17,collopy18}, and coherent control of internal molecular states~\cite{Ospelkaus2010hfcontrol, Park2017NaK, williams18,Seeselberg2018rotcoh}.

Beyond controlling the internal states and the motion of molecules, an important research frontier is to understand and control how they interact. Along this front, how molecules collide at short range is of central importance. Favorable collisional properties, i.e. high elastic scattering rate compared to inelastic loss rate~\cite{Stuhl2012evapOH,Son2019NaLiEvap}, could lead to direct evaporative cooling of molecular gases to quantum degeneracy. Full understanding of molecular collisions could also allow one to tune the elastic~\cite{Bohn2002rotFB,Lassabliere2018scatlength} and inelastic scattering rates~\cite{Gorshkov2008shield,Quemener2016shieldE,Gonzalez-Martinez2017adimtheory,Karman2018mwshield} by applying external fields. Furthermore, by elucidating the role of specific quantum states and quantum statistics of the reactants~\cite{Krems08chemistry,Carr2009review,Miranda2011stereodynamics}, collisions in the ultracold regime provide important insights into chemical and inelastic processes. For example, recent experiments have provided evidence for sticky collisions and long-lived complexes for both reactive and non-reactive molecules~\cite{Ye2018NaRb,Gregory2019RbCsSticky,Hu2019KRbReaction}.

Experimentally, inelastic collisional (loss) rates have been measured in many bi-alkali molecules either in the $^1\Sigma$ ground state or the metastable $^3\Sigma$ state, including $^{40}$K$^{87}$Rb~\cite{Ni2008KRb}, $^{23}$Na$^{40}$K~\cite{Park2015NaK}, $^{6}$Li$^{23}$Na~\cite{Rvachov2017LiNa}, $^{23}$Na$^{87}$Rb~\cite{Ye2018NaRb}, $^{87}$Rb$^{133}$Cs~\cite{Takekoshi2014RbCs,Gregory2016RbCs,Gregory2019RbCsSticky}, and $^{87}$Rb$_2$~\cite{Drews2017Rb2}. Regardless of the chemical reactivity of these molecules, all have reported loss rates within a factor of unity of that predicted by universal loss models, where two molecules have unit probability of being lost when they reach short range. This is despite the fact that some of these systems are not chemically reactive. In those systems, the loss has been interpreted as evidence for ``sticky collisions'', where a dense spectrum of molecular resonances leads to enhanced losses that approach the universal rate~\cite{Mayle2012statscat,Mayle2013resscat}. These observations illustrate that even collisions between simple, non-reactive diatomic molecules can exhibit qualitatively new features not found in atomic collisions.

Recently, direct laser-cooling and trapping of molecules~\cite{barry14,norrgard16RF,truppe17,anderegg17,collopy18} have opened the door to studying $^2\Sigma$ molecules \cite{anderegg17,williams18,McCarron18,anderegg18,cheuk18,Anderegg2019tweezer} in the ultracold regime. Compared to ground state bi-alkali molecules, the unpaired electron spin in $^2\Sigma$ molecules leads to additional features such as hyperfine structure, spin-rotational structure and intermolecular electronic spin-spin interactions at long ranges. In a previous study of bulk samples of rotationally excited $^2\Sigma$ CaF molecules, collisional loss rates near the Langevin limit were observed~\cite{Anderegg2019tweezer}. However, the method used could neither fully control the internal state nor the exact number of molecules. One could therefore not easily distinguish between various loss mechanisms such as hyperfine and rotational relaxation.

In this work, we have developed an optical tweezer based approach for studying collisions. This provides full control of the number of molecules and their internal state, overcoming previous limitations. Specifically, we prepare two single molecules in separate optical tweezer traps, prepare them in a single quantum state, and then merge them into a single trap, thereby creating an exact two-body collisional system. Our approach ensures, by construction, that only single-body and two-body processes are at play, in contrast to previous measurements in bulk molecular gases. We note that with atoms, similar tweezer-based approaches have been used to probe hyperfine relaxation~\cite{Xu2015atomtweezer,Liu2018tweezer}, coherent two-body dynamics~\cite{Sompet2019coherent}, and Feshbach resonances~\cite{Hood2019FBspec}.

\begin{figure}[t]
\centering
\includegraphics[width=87mm]{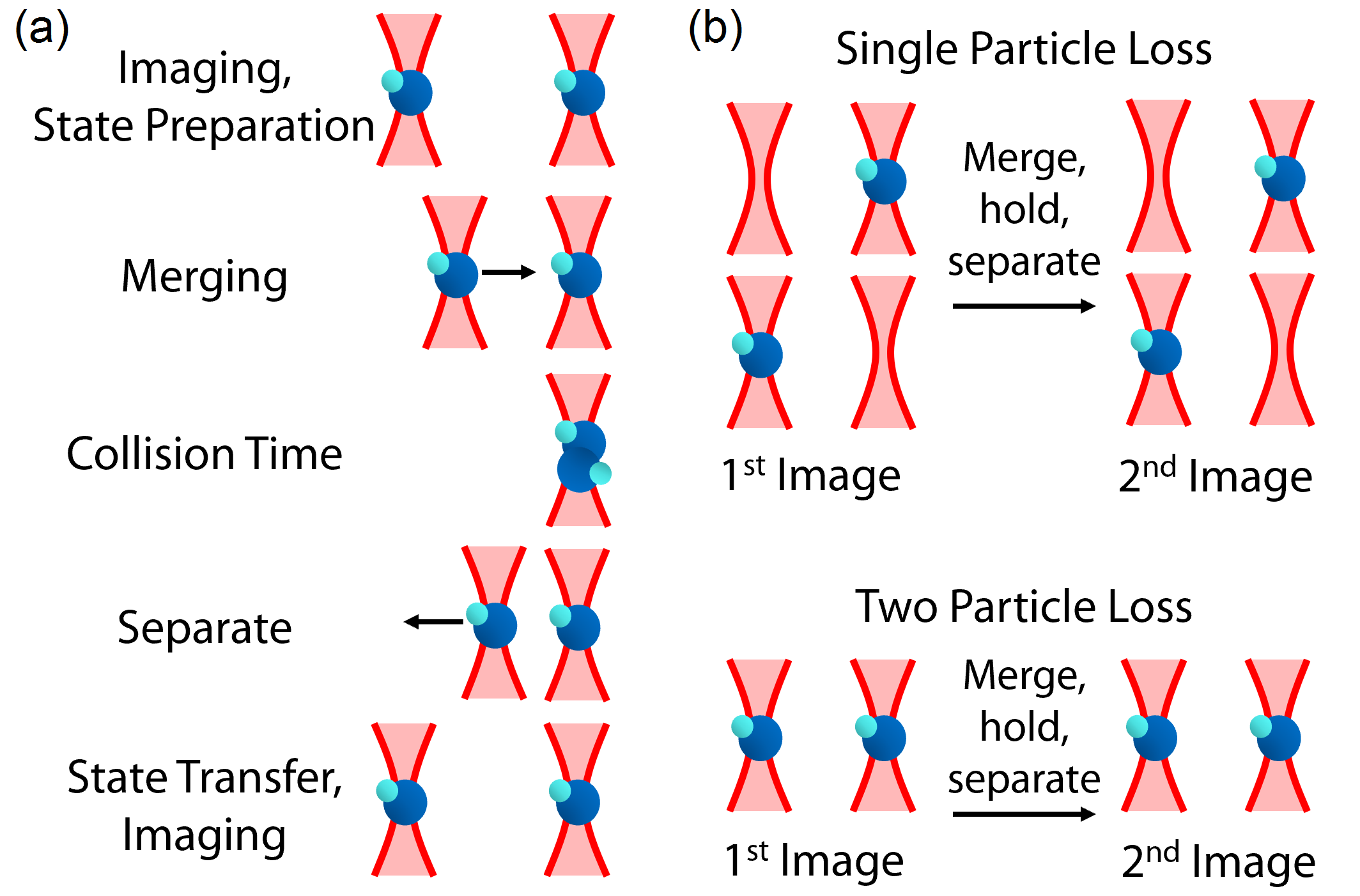}
\caption{Measuring Collisional Loss with Two Molecules. (a) Measurement protocol. Before merging and state preparation, the molecules are in $|N=1\rangle$, and a non-destructive image is taken to determine occupation of the two tweezers. After the collisional hold time, the tweezers are separated and the remaining molecules transferred back to $|N=1\rangle$, a second image is taken. (b) Determining one-particle versus two-particle loss. To measure single-particle loss, we post-select for data where one molecule is detected in the first image, and measure the survival probability in the second image. Note that the merging and splitting process randomizes the position of the molecule. To measure two-particle loss, we post-select for data where two molecules are detected, and measure the survival probability of both molecules in the second image.
}
\end{figure}

The starting point of our experiment is a magneto-optical trap (MOT) of $^{40}$Ca$^{19}$F molecules, which are in a mixture of hyperfine states in the first excited rotational manifold ($|X, N=1\rangle$). These molecules are then loaded with $\Lambda$-cooling into an optical dipole trap (ODT) formed by a single focused beam of 1064$\,$nm light. Subsequently, single molecules are stochastically loaded into two optical tweezer traps in the presence of laser cooling light~\cite{Anderegg2019tweezer}. Both tweezers have a spot size of $\sim 2\,\mu\text{m}$, and are operated at a trap depth of $1400\,\mu$K (see Supp. Mat.). The position of the right tweezer trap is fixed, while the position of the left tweezer trap can be varied. The two tweezer traps are both derived from a common laser source at 780\,nm, which is subsequently split and recombined. One path passes through an acousto-optical deflector (AOD), where the radio-frequency driving the AOD controls the position of the tweezer trap. It is then combined with the other path, generating a stationary optical tweezer trap. The $\sim 100\,\text{MHz}$ frequency difference between the beams and their orthogonal linear polarizations eliminate parametric heating that could otherwise arise due to interference. 

To study collisions, we overlap the two tweezer traps, which are each filled with at most one molecule. Since the loading is stochastic, i.e. each trap can be empty or filled,  four possible initial configurations can be realized. To identify the initial configuration, we measure the individual trap occupations using a $30\,\text{ms}$ pulse of $\Lambda$-imaging light. After detection, the molecules are brought into the desired internal state using optical and microwave pulses. The two tweezer traps are then overlapped by sweeping the left tweezer trap onto the stationary tweezer (right) over $1\,\text{ms}$. The movable trap laser beam is subsequently ramped down in $1\,\text{ms}$, slow enough to ensure adiabaticity. Following this merging step, both molecules are held together by the resultant stationary tweezer trap on the right. The two molecules are held together for a variable period of time, during which collisions can occur. The merging process is then reversed, i.e. ``splitting'', creating two tweezer traps. The surviving molecules are then transferred back into the $|N=1\rangle$ state and detected. The splitting step is necessary since during detection, if two molecules are within a single trap, rapid light-induced collisions occur and both molecules are lost. A final image of the molecules is then taken and the final occupations of the two tweezers are measured. The intensities of the two tweezer traps are tuned such that during splitting, a molecule has equal probability of ending up in either trap. If both molecules survive, they have a $50\%$ chance of being in separate traps. By post-selecting on the molecule number in the first image, we can extract both the single-particle and two-particle loss rates. In detail, single-particle loss is observed by looking at data where one molecule is detected in the first image, and measuring the probability of detecting one molecule subsequently. Two-particle loss is observed by looking at data where two molecules are loaded, and measuring the probability of detecting both molecules subsequently. By observing the single-particle and two-particle survival probabilities versus the hold time, one obtains the corresponding loss rates. If the two-particle rate is higher than the one-particle loss rate, two-body collisional loss has occurred. This measurement protocol for probing collisional loss is summarized in Fig.~1(b).

\begin{figure}[t]
\centering
\includegraphics[width=87mm]{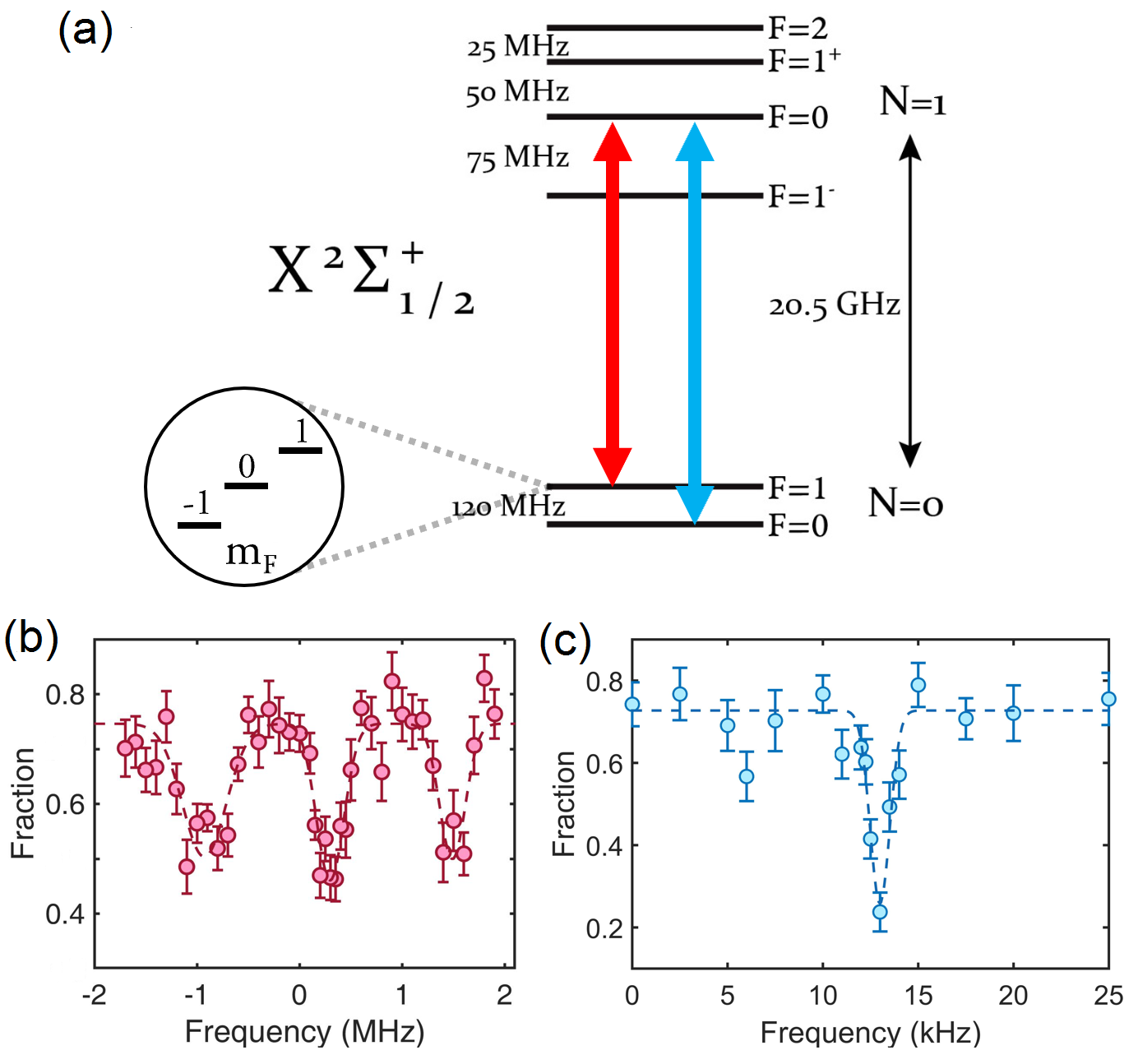} 
\caption{Preparing Molecules in the Ground Rotational Manifold. (a) Rotational and hyperfine structure in the ground electronic and vibrational state of CaF $(X,v=0)$.  (b) Microwave spectroscopy of the $|N=0, F=1\ket$ hyperfine manifold in the presence of a magnetic field of 1\,G. Molecules are prepared in $|N=1, F=0\rangle$ via optical pumping with 80\% fidelity. A single microwave pulse is applied for {20} ms, and then the surviving number of molecules in the $|N=1\rangle$ manifold is measured. The three Zeeman levels of $m_F=-1,0,1$ are well-resolved. (c) Microwave spectroscopy of the absolute ground state $|N=0, F=0\rangle$. 
Landau-Zener sweeps are used to transfer molecules prepared in $|N=1, F=0\rangle$ to the absolute ground state $|N=0, F=0\rangle$. Microwave radiation is applied, while the magnetic field is swept linearly over a range of $2.5\,\text{G}$ (center at $3.5\,\text{G}$) in $20\,\text{ms}$. Shown is the remaining fraction of molecules in $|N=1\rangle$ versus the applied microwave frequency for a fixed magnetic field sweep.}
\end{figure}

An important feature of our method is control over the internal state of the molecules. Starting with individual molecules naturally permits number-resolved state preparation. Furthermore, it also allows optical preparation of the internal state without complications from light-induced collisional loss. In the following, we describe in detail the procedure used to bring the molecules into the rotational ground state, which consists of two hyperfine manifolds, $|N=0, F=0\rangle$ and $|N=0, F=1\rangle$ (Fig.~2(a)) and then into a single hyperfine state. We perform optical pumping in combination with microwave driving on the single molecules before merging. With single molecules trapped in separate optical tweezers, light-assisted collisions, which typically lead to rapid losses at densities required to probe collisions, are not possible. We optically pump into the $|N=1, F=0, m_F=0\rangle$ state by applying $100\,\mu\text{s}$ of resonant light on the $|X, v=0, N=1, F=2,1+,1-\rangle \rightarrow |A, v=0, N=0, J=1/2\rangle$ transitions in the presence of $v=1,2,3$ vibrational repumpers. Subsequently, a microwave sweep or $\pi$-pulse is applied to bring $|N=1, F=0, m_F=0\rangle$ molecules into the desired hyperfine state in the $|N=0\rangle$ ground rotational manifold. Although nominally forbidden, the $|N=1, F=0, m_F=0\rangle \rightarrow |N=0, F=0, m_F=0\rangle$ transition can be directly driven with microwaves when a weak magnetic field ($\sim 3\,\text{G}$) is applied. The magnetic field admixes the $|N=0, F=1, m_F=0\rangle$ state (Fig. 2(c)), making the transition allowed. Due to the low resulting Rabi frequency $\sim 2\pi \times 2\,\text{kHz}$, we opted to use a Landau-Zener sweep implemented via a magnetic field sweep for robustness. The three Zeeman states of the $|N=0, F=1\rangle$ hyperfine manifold can be resolved by applying a bias magnetic field (Fig. 2(b)). For these states, where large Rabi frequencies $\sim 2\pi \times 100\,\text{kHz}$ are available, we use a single $\pi$-pulse to transfer. We achieve transfer efficiencies of {$74\%$} into the absolute ground state ($|N=0, F=0, m_F=0\rangle$), and {$70\%$} into the excited hyperfine manifold in the rotational ground state ($|N=0, F=1\rangle$) (See Supp. Mat.). Although the transfer is imperfect, any molecules remaining in the $|N=1\rangle$ rotational manifold are removed by a pulse of resonant light on the $|N=1\rangle$ cycling transition, which heats $|N=1\rangle$ molecules out of the trap. Since state preparation occurs after the initial image but before merging of the two traps, imperfect transfer leads only to an overall reduction in the surviving fraction. It does not lead to any additional background when measuring single-particle and two-particle survival.

\begin{figure}[t]
\centering
\includegraphics[width=70mm]{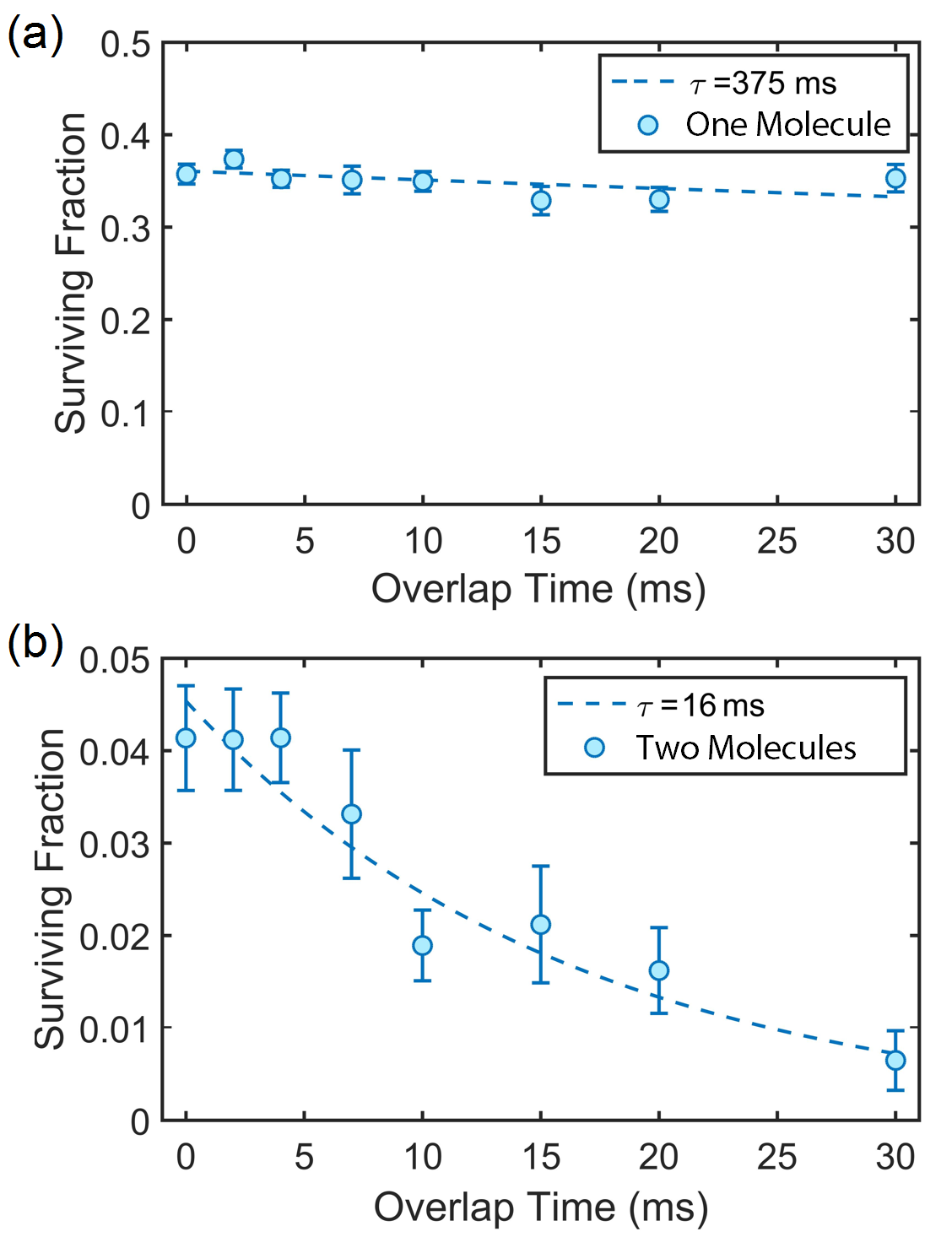}
\caption{Two-Particle Loss versus Single-Particle Loss. (a) One-particle loss measured by post-selecting on data where a single molecule is loaded. An exponential fit gives a $1/e$ time constant of $375\,\text{ms}$. (b) Two-particle loss measured by post-selecting on data where two molecules are loaded. An exponential fit gives a time constant of $16(3)\,\text{ms}$, over an order of magnitude shorter than the time constant in (a).} 
\end{figure}

For CaF molecules in the absolute ground state $|N=0, F=0\rangle$, we measure a two-particle $1/e$ lifetime of $16\,\text{ms}$, while single particle loss from the same data set yield a substantially longer lifetime of $\sim 400\,\text{ms}$ (Fig.~3(a,b)). This shows that the two particle loss is dominated by CaF-CaF two-body collisions. Since rotational relaxation and hyperfine decay cannot occur in the absolute ground state, we conclude that the loss is either due to chemical reactions ($\text{CaF}+\text{CaF} \rightarrow \text{CaF}_2+\text{Ca}$)~\cite{Meyer2011SrF,tijsemail}, or formation of complexes that are either not detectable or are lost. To convert to a two-body loss rate constant, we determine the mean density of the tweezer trap by a Monte-Carlo simulation using measured trap parameters and the molecular temperature ($41(12)\,\mu\text{K}$) via time-of-flight expansion (see Supp. Mat.). We note that the temperature achieved is lower than the $d$-wave threshold ($p$-wave and $d$-wave thresholds at $20\,\mu\text{K}$ and $106\,\mu\text{K}$ respectively). Due to the bosonic exchange statistics of the molecules when prepared in identical internal states, we expect that $s$- and $d$-wave scattering channels are active. The two-body loss rate constant for CaF molecules in the absolute ground state is found to be $7(4) \times 10^{-11}\,\text{cm}^3/\text{s}$.   

We next compare the loss rates to a single channel model~\cite{Idziaszek2010universal,Frye2015universal} with universal loss, where molecules are lost with unity probability once they approach distances much smaller than the Van der Waals length, given by $l_{vdw}=\frac{1}{2}\cos(\frac{\pi}{4}) \frac{\Gamma(3/4)}{\Gamma(5/4)} (\frac{2 \mu C_6}{\hbar^2})^{1/4} \approx 190a_0$, where $C_6$ is the van der Waals coefficient and $a_0$ is the Bohr radius. In the ground rotational state, the $C_6$ coefficient is identical for all internal states, and ignoring the small electronic contribution, is given by $C_6 = \frac{1}{(4 \pi \epsilon_0)^2} \cdot d^4/6B$, where $d$ is the dipole moment of the molecule and $B$ the rotational constant. The predicted universal loss rate, taking into account the finite temperature of $41(12)\,\mu \text{K}$, is $3.0(2) \times 10^{-10}\,\text{cm}^3/\text{s}$, which is above but close to the measured value. This is similar to measurements in bi-alkali molecules, where loss rates of the same order of magnitude as universal loss were observed~\cite{Ni2008KRb, Park2015NaK, Rvachov2017LiNa,Ye2018NaRb, Takekoshi2014RbCs,Gregory2016RbCs,Gregory2019RbCsSticky,Drews2017Rb2}.

\begin{figure}[t]
\centering
\includegraphics[width=85mm]{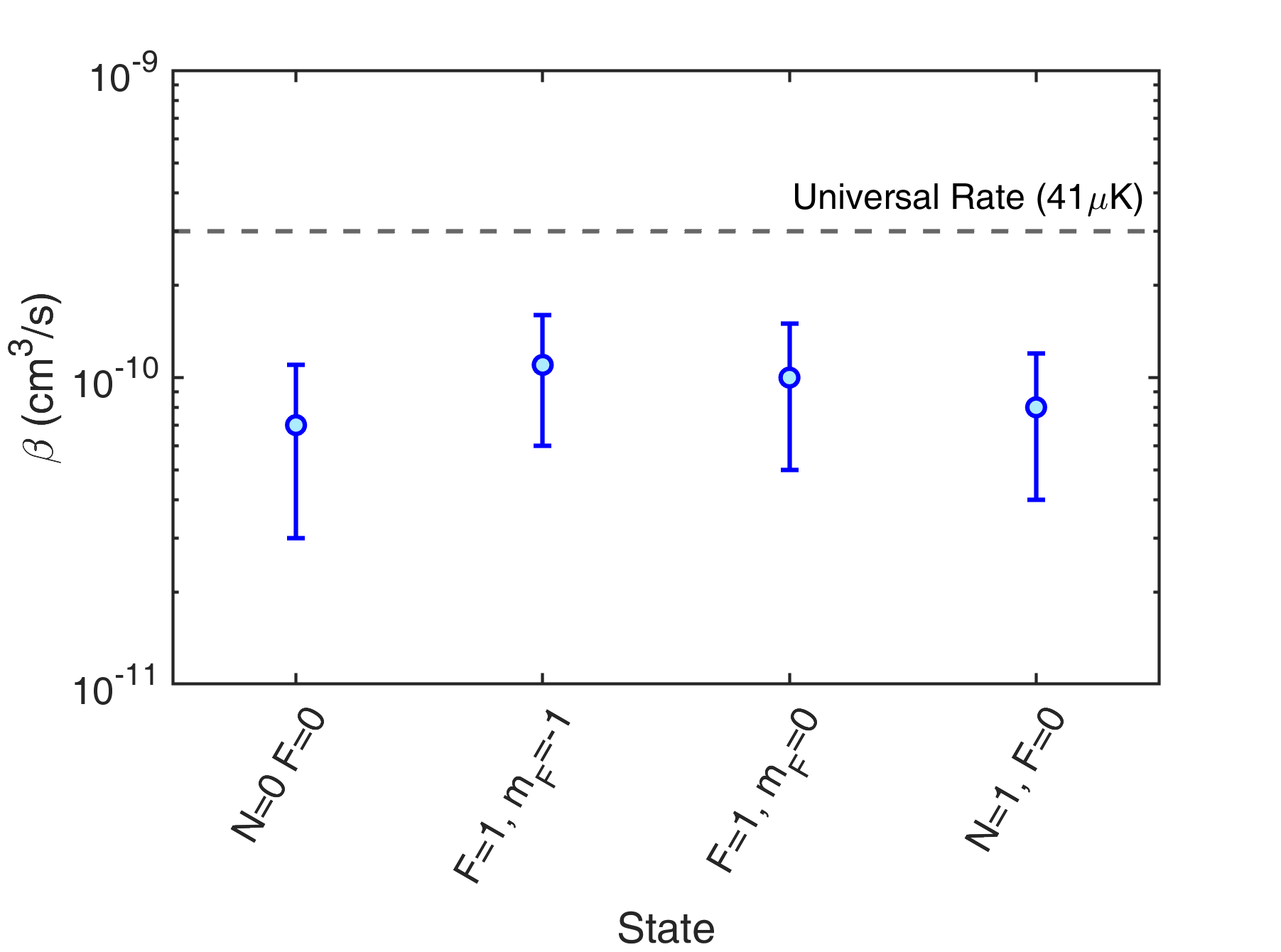}
\caption{Summary of Collisional Loss Rates. Shown are the measured collisional loss rates of various hyperfine states in the ground and first excited rotational manifold. The universal loss rate at the experimentally measured temperature of $40\,\mu\text{K}$ is shown by the dashed line. For the states $|N=0,F=1,m_F=-1,0\rangle$, the collisional rates are measured at a magnetic field of 8\,G. No significant dependence on spin polarization and hyperfine state is observed.}
\end{figure}

Since chemical reaction rates could possibly depend on the electronic spins of the $^2\Sigma$ CaF molecules, we explored the two-body loss rate of spin-polarized CaF in the ground rotational state $|N=0, F=1,m_F=-1\rangle$ at a magnetic field of 4\,G \cite{SrFChem}. We find, within the experimental uncertainty, a loss rate constant identical to molecules in the absolute ground state $|N=0, F=0, m_F=0\rangle$. This indicates that either the chemical reaction rate or complex formation rate is independent of spin. We also measure the loss rate at a higher magnetic field of 8\,G and find a similar value. We conclude that at the low magnetic fields explored, these loss mechanisms are not suppressed significantly for spin-polarized molecules. We next measure the collisional loss rate of excited hyperfine molecules in the non-spin-polarized state $|N=0, F=1, m_F=0\rangle$, which also yield a similar value. These results indicate that either complex formation or chemical reactions occur much faster than hyperfine relaxation. The loss rate measurements for various states are summarized in Fig.~4.

In conclusion, we have developed an optical tweezer-based approach to study molecular collisions, which has allowed us to measure the collisional loss rates of $^2\Sigma$ CaF molecules in their absolute rovibrational ground state, as well as selected excited hyperfine states. The measurements indicate that the dominant loss mechanism is either chemical reactions or formation of long-lived complexes. The observed loss rates do not exhibit dependence on hyperfine state, suggesting that the chemical reaction rate or the complex formation rate does not depend on the electronic spin of the molecules in the regime of magnetic fields explored. Recent theoretical proposals of trap light induced loss \cite{Christianen2019DOS} could play a role in the observed losses, but a detailed study of this is beyond the scope of the current work. In the future, collisional losses could be suppressed by implementing collisional shielding schemes that prevent molecules from reaching short range~\cite{Gorshkov2008shield,Quemener2016shieldE,Gonzalez-Martinez2017adimtheory,Karman2018mwshield}. Efficient suppression of collisional loss could open up new avenues of preparing quantum gases of molecules such as evaporative cooling~\cite{Stuhl2012evapOH,Son2019NaLiEvap} or algorithmic cooling that relies on interaction blockade~\cite{Bakr2011algo}. The tweezer-based approach developed in this work will be well-suited to exploring these possibilities. Our work can also be extended to more complex laser-coolable molecules including polyatomic ones~\cite{kozyryev16}, which could be a rich arena for explorations in ultracold collisions and reactions~\cite{Krems08chemistry, Carr2009review}, as well as quantum simulation~\cite{blackmore18}.

This work was supported by the NSF and ARO. We thank John Bohn, Tijs Karman and Goulven Quemener for insightful discussions and universal loss calculations. LWC acknowledges supports from the MPHQ. LA acknowledges support from HQI. SB acknowledges supports from the NSF GRFP.


\begin{thebibliography}{49}%
\makeatletter
\providecommand \@ifxundefined [1]{%
 \@ifx{#1\undefined}
}%
\providecommand \@ifnum [1]{%
 \ifnum #1\expandafter \@firstoftwo
 \else \expandafter \@secondoftwo
 \fi
}%
\providecommand \@ifx [1]{%
 \ifx #1\expandafter \@firstoftwo
 \else \expandafter \@secondoftwo
 \fi
}%
\providecommand \natexlab [1]{#1}%
\providecommand \enquote  [1]{``#1''}%
\providecommand \bibnamefont  [1]{#1}%
\providecommand \bibfnamefont [1]{#1}%
\providecommand \citenamefont [1]{#1}%
\providecommand \href@noop [0]{\@secondoftwo}%
\providecommand \href [0]{\begingroup \@sanitize@url \@href}%
\providecommand \@href[1]{\@@startlink{#1}\@@href}%
\providecommand \@@href[1]{\endgroup#1\@@endlink}%
\providecommand \@sanitize@url [0]{\catcode `\\12\catcode `\$12\catcode
  `\&12\catcode `\#12\catcode `\^12\catcode `\_12\catcode `\%12\relax}%
\providecommand \@@startlink[1]{}%
\providecommand \@@endlink[0]{}%
\providecommand \url  [0]{\begingroup\@sanitize@url \@url }%
\providecommand \@url [1]{\endgroup\@href {#1}{\urlprefix }}%
\providecommand \urlprefix  [0]{URL }%
\providecommand \Eprint [0]{\href }%
\providecommand \doibase [0]{http://dx.doi.org/}%
\providecommand \selectlanguage [0]{\@gobble}%
\providecommand \bibinfo  [0]{\@secondoftwo}%
\providecommand \bibfield  [0]{\@secondoftwo}%
\providecommand \translation [1]{[#1]}%
\providecommand \BibitemOpen [0]{}%
\providecommand \bibitemStop [0]{}%
\providecommand \bibitemNoStop [0]{.\EOS\space}%
\providecommand \EOS [0]{\spacefactor3000\relax}%
\providecommand \BibitemShut  [1]{\csname bibitem#1\endcsname}%
\let\auto@bib@innerbib\@empty
\bibitem [{\citenamefont {Carr}\ \emph {et~al.}(2009)\citenamefont {Carr},
  \citenamefont {DeMille}, \citenamefont {Krems},\ and\ \citenamefont
  {Ye}}]{Carr2009review}%
  \BibitemOpen
  \bibfield  {author} {\bibinfo {author} {\bibfnamefont {L.~D.}\ \bibnamefont
  {Carr}}, \bibinfo {author} {\bibfnamefont {D.}~\bibnamefont {DeMille}},
  \bibinfo {author} {\bibfnamefont {R.~V.}\ \bibnamefont {Krems}}, \ and\
  \bibinfo {author} {\bibfnamefont {J.}~\bibnamefont {Ye}},\ }\href@noop {}
  {\bibfield  {journal} {\bibinfo  {journal} {New J. Phys.}\ }\textbf {\bibinfo
  {volume} {11}} (\bibinfo {year} {2009})}\BibitemShut {NoStop}%
\bibitem [{\citenamefont {Ni}\ \emph {et~al.}(2008)\citenamefont {Ni},
  \citenamefont {Ospelkaus}, \citenamefont {de~Miranda}, \citenamefont {Pe'er},
  \citenamefont {Neyenhuis}, \citenamefont {Zirbel}, \citenamefont
  {Kotochigova}, \citenamefont {Julienne}, \citenamefont {Jin},\ and\
  \citenamefont {Ye}}]{Ni2008KRb}%
  \BibitemOpen
  \bibfield  {author} {\bibinfo {author} {\bibfnamefont {K.-K.}\ \bibnamefont
  {Ni}}, \bibinfo {author} {\bibfnamefont {S.}~\bibnamefont {Ospelkaus}},
  \bibinfo {author} {\bibfnamefont {M.~H.~G.}\ \bibnamefont {de~Miranda}},
  \bibinfo {author} {\bibfnamefont {A.}~\bibnamefont {Pe'er}}, \bibinfo
  {author} {\bibfnamefont {B.}~\bibnamefont {Neyenhuis}}, \bibinfo {author}
  {\bibfnamefont {J.~J.}\ \bibnamefont {Zirbel}}, \bibinfo {author}
  {\bibfnamefont {S.}~\bibnamefont {Kotochigova}}, \bibinfo {author}
  {\bibfnamefont {P.~S.}\ \bibnamefont {Julienne}}, \bibinfo {author}
  {\bibfnamefont {D.~S.}\ \bibnamefont {Jin}}, \ and\ \bibinfo {author}
  {\bibfnamefont {J.}~\bibnamefont {Ye}},\ }\href@noop {} {\bibfield  {journal}
  {\bibinfo  {journal} {Science}\ }\textbf {\bibinfo {volume} {322}},\ \bibinfo
  {pages} {231} (\bibinfo {year} {2008})}\BibitemShut {NoStop}%
\bibitem [{\citenamefont {Takekoshi}\ \emph {et~al.}(2014)\citenamefont
  {Takekoshi}, \citenamefont {ReichsÃ¶llner}, \citenamefont {Schindewolf},
  \citenamefont {Hutson}, \citenamefont {Le~Sueur}, \citenamefont {Dulieu},
  \citenamefont {Ferlaino}, \citenamefont {Grimm},\ and\ \citenamefont
  {NÃ¤gerl}}]{Takekoshi2014RbCs}%
  \BibitemOpen
  \bibfield  {author} {\bibinfo {author} {\bibfnamefont {T.}~\bibnamefont
  {Takekoshi}}, \bibinfo {author} {\bibfnamefont {L.}~\bibnamefont
  {ReichsÃ¶llner}}, \bibinfo {author} {\bibfnamefont {A.}~\bibnamefont
  {Schindewolf}}, \bibinfo {author} {\bibfnamefont {J.~M.}\ \bibnamefont
  {Hutson}}, \bibinfo {author} {\bibfnamefont {C.~R.}\ \bibnamefont
  {Le~Sueur}}, \bibinfo {author} {\bibfnamefont {O.}~\bibnamefont {Dulieu}},
  \bibinfo {author} {\bibfnamefont {F.}~\bibnamefont {Ferlaino}}, \bibinfo
  {author} {\bibfnamefont {R.}~\bibnamefont {Grimm}}, \ and\ \bibinfo {author}
  {\bibfnamefont {H.-C.}\ \bibnamefont {NÃ¤gerl}},\ }\href
  {https://link.aps.org/doi/10.1103/PhysRevLett.113.205301} {\bibfield
  {journal} {\bibinfo  {journal} {Phys. Rev. Lett.}\ }\textbf {\bibinfo
  {volume} {113}},\ \bibinfo {pages} {205301} (\bibinfo {year}
  {2014})}\BibitemShut {NoStop}%
\bibitem [{\citenamefont {Park}\ \emph {et~al.}(2015)\citenamefont {Park},
  \citenamefont {Will},\ and\ \citenamefont {Zwierlein}}]{Park2015NaK}%
  \BibitemOpen
  \bibfield  {author} {\bibinfo {author} {\bibfnamefont {J.~W.}\ \bibnamefont
  {Park}}, \bibinfo {author} {\bibfnamefont {S.~A.}\ \bibnamefont {Will}}, \
  and\ \bibinfo {author} {\bibfnamefont {M.~W.}\ \bibnamefont {Zwierlein}},\
  }\href {https://link.aps.org/doi/10.1103/PhysRevLett.114.205302} {\bibfield
  {journal} {\bibinfo  {journal} {Phys. Rev. Lett.}\ }\textbf {\bibinfo
  {volume} {114}},\ \bibinfo {pages} {205302} (\bibinfo {year}
  {2015})}\BibitemShut {NoStop}%
\bibitem [{\citenamefont {Gregory}\ \emph {et~al.}(2016)\citenamefont
  {Gregory}, \citenamefont {Aldegunde}, \citenamefont {Hutson},\ and\
  \citenamefont {Cornish}}]{Gregory2016RbCs}%
  \BibitemOpen
  \bibfield  {author} {\bibinfo {author} {\bibfnamefont {P.~D.}\ \bibnamefont
  {Gregory}}, \bibinfo {author} {\bibfnamefont {J.}~\bibnamefont {Aldegunde}},
  \bibinfo {author} {\bibfnamefont {J.~M.}\ \bibnamefont {Hutson}}, \ and\
  \bibinfo {author} {\bibfnamefont {S.~L.}\ \bibnamefont {Cornish}},\ }\href
  {https://link.aps.org/doi/10.1103/PhysRevA.94.041403} {\bibfield  {journal}
  {\bibinfo  {journal} {Phys. Rev. A}\ }\textbf {\bibinfo {volume} {94}},\
  \bibinfo {pages} {041403} (\bibinfo {year} {2016})}\BibitemShut {NoStop}%
\bibitem [{\citenamefont {Guo}\ \emph {et~al.}(2017)\citenamefont {Guo},
  \citenamefont {Vexiau}, \citenamefont {Zhu}, \citenamefont {Lu},
  \citenamefont {Bouloufa-Maafa}, \citenamefont {Dulieu},\ and\ \citenamefont
  {Wang}}]{Guo2017NaRb}%
  \BibitemOpen
  \bibfield  {author} {\bibinfo {author} {\bibfnamefont {M.}~\bibnamefont
  {Guo}}, \bibinfo {author} {\bibfnamefont {R.}~\bibnamefont {Vexiau}},
  \bibinfo {author} {\bibfnamefont {B.}~\bibnamefont {Zhu}}, \bibinfo {author}
  {\bibfnamefont {B.}~\bibnamefont {Lu}}, \bibinfo {author} {\bibfnamefont
  {N.}~\bibnamefont {Bouloufa-Maafa}}, \bibinfo {author} {\bibfnamefont
  {O.}~\bibnamefont {Dulieu}}, \ and\ \bibinfo {author} {\bibfnamefont
  {D.}~\bibnamefont {Wang}},\ }\href
  {https://link.aps.org/doi/10.1103/PhysRevA.96.052505} {\bibfield  {journal}
  {\bibinfo  {journal} {Phys. Rev. A}\ }\textbf {\bibinfo {volume} {96}},\
  \bibinfo {pages} {052505} (\bibinfo {year} {2017})}\BibitemShut {NoStop}%
\bibitem [{\citenamefont {Rvachov}\ \emph {et~al.}(2017)\citenamefont
  {Rvachov}, \citenamefont {Son}, \citenamefont {Sommer}, \citenamefont
  {Ebadi}, \citenamefont {Park}, \citenamefont {Zwierlein}, \citenamefont
  {Ketterle},\ and\ \citenamefont {Jamison}}]{Rvachov2017LiNa}%
  \BibitemOpen
  \bibfield  {author} {\bibinfo {author} {\bibfnamefont {T.~M.}\ \bibnamefont
  {Rvachov}}, \bibinfo {author} {\bibfnamefont {H.}~\bibnamefont {Son}},
  \bibinfo {author} {\bibfnamefont {A.~T.}\ \bibnamefont {Sommer}}, \bibinfo
  {author} {\bibfnamefont {S.}~\bibnamefont {Ebadi}}, \bibinfo {author}
  {\bibfnamefont {J.~J.}\ \bibnamefont {Park}}, \bibinfo {author}
  {\bibfnamefont {M.~W.}\ \bibnamefont {Zwierlein}}, \bibinfo {author}
  {\bibfnamefont {W.}~\bibnamefont {Ketterle}}, \ and\ \bibinfo {author}
  {\bibfnamefont {A.~O.}\ \bibnamefont {Jamison}},\ }\href
  {https://link.aps.org/doi/10.1103/PhysRevLett.119.143001} {\bibfield
  {journal} {\bibinfo  {journal} {Phys. Rev. Lett.}\ }\textbf {\bibinfo
  {volume} {119}},\ \bibinfo {pages} {143001} (\bibinfo {year}
  {2017})}\BibitemShut {NoStop}%
\bibitem [{\citenamefont {Barry}\ \emph {et~al.}(2014)\citenamefont {Barry},
  \citenamefont {McCarron}, \citenamefont {Norrgard}, \citenamefont
  {Steinecker},\ and\ \citenamefont {DeMille}}]{barry14}%
  \BibitemOpen
  \bibfield  {author} {\bibinfo {author} {\bibfnamefont {J.~F.}\ \bibnamefont
  {Barry}}, \bibinfo {author} {\bibfnamefont {D.~J.}\ \bibnamefont {McCarron}},
  \bibinfo {author} {\bibfnamefont {E.~B.}\ \bibnamefont {Norrgard}}, \bibinfo
  {author} {\bibfnamefont {M.~H.}\ \bibnamefont {Steinecker}}, \ and\ \bibinfo
  {author} {\bibfnamefont {D.}~\bibnamefont {DeMille}},\ }\href {\doibase
  10.1038/nature13634} {\bibfield  {journal} {\bibinfo  {journal} {Nature}\
  }\textbf {\bibinfo {volume} {512}},\ \bibinfo {pages} {286} (\bibinfo {year}
  {2014})}\BibitemShut {NoStop}%
\bibitem [{\citenamefont {Truppe}\ \emph {et~al.}(2017)\citenamefont {Truppe},
  \citenamefont {Williams}, \citenamefont {Hambach}, \citenamefont {Caldwell},
  \citenamefont {Fitch}, \citenamefont {Hinds}, \citenamefont {Sauer},\ and\
  \citenamefont {Tarbutt}}]{truppe17}%
  \BibitemOpen
  \bibfield  {author} {\bibinfo {author} {\bibfnamefont {S.}~\bibnamefont
  {Truppe}}, \bibinfo {author} {\bibfnamefont {H.~J.}\ \bibnamefont
  {Williams}}, \bibinfo {author} {\bibfnamefont {M.}~\bibnamefont {Hambach}},
  \bibinfo {author} {\bibfnamefont {L.}~\bibnamefont {Caldwell}}, \bibinfo
  {author} {\bibfnamefont {N.~J.}\ \bibnamefont {Fitch}}, \bibinfo {author}
  {\bibfnamefont {E.~A.}\ \bibnamefont {Hinds}}, \bibinfo {author}
  {\bibfnamefont {B.~E.}\ \bibnamefont {Sauer}}, \ and\ \bibinfo {author}
  {\bibfnamefont {M.~R.}\ \bibnamefont {Tarbutt}},\ }\href {\doibase
  10.1038/nphys4241} {\bibfield  {journal} {\bibinfo  {journal} {Nature
  Physics}\ }\textbf {\bibinfo {volume} {13}},\ \bibinfo {pages} {1173}
  (\bibinfo {year} {2017})}\BibitemShut {NoStop}%
\bibitem [{\citenamefont {Anderegg}\ \emph {et~al.}(2017)\citenamefont
  {Anderegg}, \citenamefont {Augenbraun}, \citenamefont {Chae}, \citenamefont
  {Hemmerling}, \citenamefont {Hutzler}, \citenamefont {Ravi}, \citenamefont
  {Collopy}, \citenamefont {Ye}, \citenamefont {Ketterle},\ and\ \citenamefont
  {Doyle}}]{anderegg17}%
  \BibitemOpen
  \bibfield  {author} {\bibinfo {author} {\bibfnamefont {L.}~\bibnamefont
  {Anderegg}}, \bibinfo {author} {\bibfnamefont {B.~L.}\ \bibnamefont
  {Augenbraun}}, \bibinfo {author} {\bibfnamefont {E.}~\bibnamefont {Chae}},
  \bibinfo {author} {\bibfnamefont {B.}~\bibnamefont {Hemmerling}}, \bibinfo
  {author} {\bibfnamefont {N.~R.}\ \bibnamefont {Hutzler}}, \bibinfo {author}
  {\bibfnamefont {A.}~\bibnamefont {Ravi}}, \bibinfo {author} {\bibfnamefont
  {A.}~\bibnamefont {Collopy}}, \bibinfo {author} {\bibfnamefont
  {J.}~\bibnamefont {Ye}}, \bibinfo {author} {\bibfnamefont {W.}~\bibnamefont
  {Ketterle}}, \ and\ \bibinfo {author} {\bibfnamefont {J.~M.}\ \bibnamefont
  {Doyle}},\ }\href@noop {} {\bibfield  {journal} {\bibinfo  {journal}
  {Physical Review Letters}\ }\textbf {\bibinfo {volume} {119}} (\bibinfo
  {year} {2017})}\BibitemShut {NoStop}%
\bibitem [{\citenamefont {Collopy}\ \emph {et~al.}(2018)\citenamefont
  {Collopy}, \citenamefont {Ding}, \citenamefont {Wu}, \citenamefont
  {Finneran}, \citenamefont {Anderegg}, \citenamefont {Augenbraun},
  \citenamefont {Doyle},\ and\ \citenamefont {Ye}}]{collopy18}%
  \BibitemOpen
  \bibfield  {author} {\bibinfo {author} {\bibfnamefont {A.~L.}\ \bibnamefont
  {Collopy}}, \bibinfo {author} {\bibfnamefont {S.}~\bibnamefont {Ding}},
  \bibinfo {author} {\bibfnamefont {Y.}~\bibnamefont {Wu}}, \bibinfo {author}
  {\bibfnamefont {I.~A.}\ \bibnamefont {Finneran}}, \bibinfo {author}
  {\bibfnamefont {L.}~\bibnamefont {Anderegg}}, \bibinfo {author}
  {\bibfnamefont {B.~L.}\ \bibnamefont {Augenbraun}}, \bibinfo {author}
  {\bibfnamefont {J.~M.}\ \bibnamefont {Doyle}}, \ and\ \bibinfo {author}
  {\bibfnamefont {J.}~\bibnamefont {Ye}},\ }\href@noop {} {\bibfield  {journal}
  {\bibinfo  {journal} {Physical Review Letters}\ }\textbf {\bibinfo {volume}
  {121}} (\bibinfo {year} {2018})}\BibitemShut {NoStop}%
\bibitem [{\citenamefont {Ospelkaus}\ \emph {et~al.}(2010)\citenamefont
  {Ospelkaus}, \citenamefont {Ni}, \citenamefont {Qu\'em\'ener}, \citenamefont
  {Neyenhuis}, \citenamefont {Wang}, \citenamefont {de~Miranda}, \citenamefont
  {Bohn}, \citenamefont {Ye},\ and\ \citenamefont
  {Jin}}]{Ospelkaus2010hfcontrol}%
  \BibitemOpen
  \bibfield  {author} {\bibinfo {author} {\bibfnamefont {S.}~\bibnamefont
  {Ospelkaus}}, \bibinfo {author} {\bibfnamefont {K.-K.}\ \bibnamefont {Ni}},
  \bibinfo {author} {\bibfnamefont {G.}~\bibnamefont {Qu\'em\'ener}}, \bibinfo
  {author} {\bibfnamefont {B.}~\bibnamefont {Neyenhuis}}, \bibinfo {author}
  {\bibfnamefont {D.}~\bibnamefont {Wang}}, \bibinfo {author} {\bibfnamefont
  {M.~H.~G.}\ \bibnamefont {de~Miranda}}, \bibinfo {author} {\bibfnamefont
  {J.~L.}\ \bibnamefont {Bohn}}, \bibinfo {author} {\bibfnamefont
  {J.}~\bibnamefont {Ye}}, \ and\ \bibinfo {author} {\bibfnamefont {D.~S.}\
  \bibnamefont {Jin}},\ }\href {\doibase 10.1103/PhysRevLett.104.030402}
  {\bibfield  {journal} {\bibinfo  {journal} {Phys. Rev. Lett.}\ }\textbf
  {\bibinfo {volume} {104}},\ \bibinfo {pages} {030402} (\bibinfo {year}
  {2010})}\BibitemShut {NoStop}%
\bibitem [{\citenamefont {Park}\ \emph {et~al.}(2017)\citenamefont {Park},
  \citenamefont {Yan}, \citenamefont {Loh}, \citenamefont {Will},\ and\
  \citenamefont {Zwierlein}}]{Park2017NaK}%
  \BibitemOpen
  \bibfield  {author} {\bibinfo {author} {\bibfnamefont {J.~W.}\ \bibnamefont
  {Park}}, \bibinfo {author} {\bibfnamefont {Z.~Z.}\ \bibnamefont {Yan}},
  \bibinfo {author} {\bibfnamefont {H.}~\bibnamefont {Loh}}, \bibinfo {author}
  {\bibfnamefont {S.~A.}\ \bibnamefont {Will}}, \ and\ \bibinfo {author}
  {\bibfnamefont {M.~W.}\ \bibnamefont {Zwierlein}},\ }\href {\doibase
  10.1126/science.aal5066} {\bibfield  {journal} {\bibinfo  {journal}
  {Science}\ }\textbf {\bibinfo {volume} {357}},\ \bibinfo {pages} {372}
  (\bibinfo {year} {2017})}\BibitemShut {NoStop}%
\bibitem [{\citenamefont {Williams}\ \emph {et~al.}(2018)\citenamefont
  {Williams}, \citenamefont {Caldwell}, \citenamefont {Fitch}, \citenamefont
  {Truppe}, \citenamefont {Rodewald}, \citenamefont {Hinds}, \citenamefont
  {Sauer},\ and\ \citenamefont {Tarbutt}}]{williams18}%
  \BibitemOpen
  \bibfield  {author} {\bibinfo {author} {\bibfnamefont {H.}~\bibnamefont
  {Williams}}, \bibinfo {author} {\bibfnamefont {L.}~\bibnamefont {Caldwell}},
  \bibinfo {author} {\bibfnamefont {N.}~\bibnamefont {Fitch}}, \bibinfo
  {author} {\bibfnamefont {S.}~\bibnamefont {Truppe}}, \bibinfo {author}
  {\bibfnamefont {J.}~\bibnamefont {Rodewald}}, \bibinfo {author}
  {\bibfnamefont {E.}~\bibnamefont {Hinds}}, \bibinfo {author} {\bibfnamefont
  {B.}~\bibnamefont {Sauer}}, \ and\ \bibinfo {author} {\bibfnamefont
  {M.}~\bibnamefont {Tarbutt}},\ }\href@noop {} {\bibfield  {journal} {\bibinfo
   {journal} {Physical Review Letters}\ }\textbf {\bibinfo {volume} {120}}
  (\bibinfo {year} {2018})}\BibitemShut {NoStop}%
\bibitem [{\citenamefont {See\ss{}elberg}\ \emph {et~al.}(2018)\citenamefont
  {See\ss{}elberg}, \citenamefont {Luo}, \citenamefont {Li}, \citenamefont
  {Bause}, \citenamefont {Kotochigova}, \citenamefont {Bloch},\ and\
  \citenamefont {Gohle}}]{Seeselberg2018rotcoh}%
  \BibitemOpen
  \bibfield  {author} {\bibinfo {author} {\bibfnamefont {F.}~\bibnamefont
  {See\ss{}elberg}}, \bibinfo {author} {\bibfnamefont {X.-Y.}\ \bibnamefont
  {Luo}}, \bibinfo {author} {\bibfnamefont {M.}~\bibnamefont {Li}}, \bibinfo
  {author} {\bibfnamefont {R.}~\bibnamefont {Bause}}, \bibinfo {author}
  {\bibfnamefont {S.}~\bibnamefont {Kotochigova}}, \bibinfo {author}
  {\bibfnamefont {I.}~\bibnamefont {Bloch}}, \ and\ \bibinfo {author}
  {\bibfnamefont {C.}~\bibnamefont {Gohle}},\ }\href {\doibase
  10.1103/PhysRevLett.121.253401} {\bibfield  {journal} {\bibinfo  {journal}
  {Phys. Rev. Lett.}\ }\textbf {\bibinfo {volume} {121}},\ \bibinfo {pages}
  {253401} (\bibinfo {year} {2018})}\BibitemShut {NoStop}%
\bibitem [{\citenamefont {Stuhl}\ \emph {et~al.}(2012)\citenamefont {Stuhl},
  \citenamefont {Hummon}, \citenamefont {Yeo}, \citenamefont {QuÃ©mÃ©ner},
  \citenamefont {Bohn},\ and\ \citenamefont {Ye}}]{Stuhl2012evapOH}%
  \BibitemOpen
  \bibfield  {author} {\bibinfo {author} {\bibfnamefont {B.~K.}\ \bibnamefont
  {Stuhl}}, \bibinfo {author} {\bibfnamefont {M.~T.}\ \bibnamefont {Hummon}},
  \bibinfo {author} {\bibfnamefont {M.}~\bibnamefont {Yeo}}, \bibinfo {author}
  {\bibfnamefont {G.}~\bibnamefont {QuÃ©mÃ©ner}}, \bibinfo {author}
  {\bibfnamefont {J.~L.}\ \bibnamefont {Bohn}}, \ and\ \bibinfo {author}
  {\bibfnamefont {J.}~\bibnamefont {Ye}},\ }\href
  {https://doi.org/10.1038/nature11718} {\bibfield  {journal} {\bibinfo
  {journal} {Nature}\ }\textbf {\bibinfo {volume} {492}},\ \bibinfo {pages}
  {396} (\bibinfo {year} {2012})}\BibitemShut {NoStop}%
\bibitem [{\citenamefont {Son}\ \emph {et~al.}(2019)\citenamefont {Son},
  \citenamefont {Park}, \citenamefont {Ketterle},\ and\ \citenamefont
  {Jamison}}]{Son2019NaLiEvap}%
  \BibitemOpen
  \bibfield  {author} {\bibinfo {author} {\bibfnamefont {H.}~\bibnamefont
  {Son}}, \bibinfo {author} {\bibfnamefont {J.~J.}\ \bibnamefont {Park}},
  \bibinfo {author} {\bibfnamefont {W.}~\bibnamefont {Ketterle}}, \ and\
  \bibinfo {author} {\bibfnamefont {A.~O.}\ \bibnamefont {Jamison}},\
  }\href@noop {} {\bibfield  {journal} {\bibinfo  {journal}
  {arXiv:1907.09630v1}\ } (\bibinfo {year} {2019})}\BibitemShut {NoStop}%
\bibitem [{\citenamefont {Bohn}\ \emph {et~al.}(2002)\citenamefont {Bohn},
  \citenamefont {Avdeenkov},\ and\ \citenamefont {Deskevich}}]{Bohn2002rotFB}%
  \BibitemOpen
  \bibfield  {author} {\bibinfo {author} {\bibfnamefont {J.~L.}\ \bibnamefont
  {Bohn}}, \bibinfo {author} {\bibfnamefont {A.~V.}\ \bibnamefont {Avdeenkov}},
  \ and\ \bibinfo {author} {\bibfnamefont {M.~P.}\ \bibnamefont {Deskevich}},\
  }\href {https://link.aps.org/doi/10.1103/PhysRevLett.89.203202} {\bibfield
  {journal} {\bibinfo  {journal} {Phys. Rev. Lett.}\ }\textbf {\bibinfo
  {volume} {89}},\ \bibinfo {pages} {203202} (\bibinfo {year}
  {2002})}\BibitemShut {NoStop}%
\bibitem [{\citenamefont {LassabliÃ¨re}\ and\ \citenamefont
  {QuÃ©mÃ©ner}(2018)}]{Lassabliere2018scatlength}%
  \BibitemOpen
  \bibfield  {author} {\bibinfo {author} {\bibfnamefont {L.}~\bibnamefont
  {LassabliÃ¨re}}\ and\ \bibinfo {author} {\bibfnamefont {G.}~\bibnamefont
  {QuÃ©mÃ©ner}},\ }\href
  {https://link.aps.org/doi/10.1103/PhysRevLett.121.163402} {\bibfield
  {journal} {\bibinfo  {journal} {Phys. Rev. Lett.}\ }\textbf {\bibinfo
  {volume} {121}},\ \bibinfo {pages} {163402} (\bibinfo {year}
  {2018})}\BibitemShut {NoStop}%
\bibitem [{\citenamefont {Gorshkov}\ \emph {et~al.}(2008)\citenamefont
  {Gorshkov}, \citenamefont {Rabl}, \citenamefont {Pupillo}, \citenamefont
  {Micheli}, \citenamefont {Zoller}, \citenamefont {Lukin},\ and\ \citenamefont
  {B\"uchler}}]{Gorshkov2008shield}%
  \BibitemOpen
  \bibfield  {author} {\bibinfo {author} {\bibfnamefont {A.~V.}\ \bibnamefont
  {Gorshkov}}, \bibinfo {author} {\bibfnamefont {P.}~\bibnamefont {Rabl}},
  \bibinfo {author} {\bibfnamefont {G.}~\bibnamefont {Pupillo}}, \bibinfo
  {author} {\bibfnamefont {A.}~\bibnamefont {Micheli}}, \bibinfo {author}
  {\bibfnamefont {P.}~\bibnamefont {Zoller}}, \bibinfo {author} {\bibfnamefont
  {M.~D.}\ \bibnamefont {Lukin}}, \ and\ \bibinfo {author} {\bibfnamefont
  {H.~P.}\ \bibnamefont {B\"uchler}},\ }\href {\doibase
  10.1103/PhysRevLett.101.073201} {\bibfield  {journal} {\bibinfo  {journal}
  {Phys. Rev. Lett.}\ }\textbf {\bibinfo {volume} {101}},\ \bibinfo {pages}
  {073201} (\bibinfo {year} {2008})}\BibitemShut {NoStop}%
\bibitem [{\citenamefont {Qu\'em\'ener}\ and\ \citenamefont
  {Bohn}(2016)}]{Quemener2016shieldE}%
  \BibitemOpen
  \bibfield  {author} {\bibinfo {author} {\bibfnamefont {G.}~\bibnamefont
  {Qu\'em\'ener}}\ and\ \bibinfo {author} {\bibfnamefont {J.~L.}\ \bibnamefont
  {Bohn}},\ }\href {\doibase 10.1103/PhysRevA.93.012704} {\bibfield  {journal}
  {\bibinfo  {journal} {Phys. Rev. A}\ }\textbf {\bibinfo {volume} {93}},\
  \bibinfo {pages} {012704} (\bibinfo {year} {2016})}\BibitemShut {NoStop}%
\bibitem [{\citenamefont {GonzÃ¡lez-MartÃ­nez}\ \emph
  {et~al.}(2017)\citenamefont {GonzÃ¡lez-MartÃ­nez}, \citenamefont {Bohn},\
  and\ \citenamefont {QuÃ©mÃ©ner}}]{Gonzalez-Martinez2017adimtheory}%
  \BibitemOpen
  \bibfield  {author} {\bibinfo {author} {\bibfnamefont {M.~L.}\ \bibnamefont
  {GonzÃ¡lez-MartÃ­nez}}, \bibinfo {author} {\bibfnamefont {J.~L.}\
  \bibnamefont {Bohn}}, \ and\ \bibinfo {author} {\bibfnamefont
  {G.}~\bibnamefont {QuÃ©mÃ©ner}},\ }\href
  {https://link.aps.org/doi/10.1103/PhysRevA.96.032718} {\bibfield  {journal}
  {\bibinfo  {journal} {Phys. Rev. A}\ }\textbf {\bibinfo {volume} {96}},\
  \bibinfo {pages} {032718} (\bibinfo {year} {2017})}\BibitemShut {NoStop}%
\bibitem [{\citenamefont {Karman}\ and\ \citenamefont
  {Hutson}(2018)}]{Karman2018mwshield}%
  \BibitemOpen
  \bibfield  {author} {\bibinfo {author} {\bibfnamefont {T.}~\bibnamefont
  {Karman}}\ and\ \bibinfo {author} {\bibfnamefont {J.~M.}\ \bibnamefont
  {Hutson}},\ }\href {\doibase 10.1103/PhysRevLett.121.163401} {\bibfield
  {journal} {\bibinfo  {journal} {Phys. Rev. Lett.}\ }\textbf {\bibinfo
  {volume} {121}},\ \bibinfo {pages} {163401} (\bibinfo {year}
  {2018})}\BibitemShut {NoStop}%
\bibitem [{\citenamefont {Krems}(2008)}]{Krems08chemistry}%
  \BibitemOpen
  \bibfield  {author} {\bibinfo {author} {\bibfnamefont {R.~V.}\ \bibnamefont
  {Krems}},\ }\href {\doibase 10.1039/b802322k} {\bibfield  {journal} {\bibinfo
   {journal} {Physical Chemistry Chemical Physics}\ }\textbf {\bibinfo {volume}
  {10}},\ \bibinfo {pages} {4079} (\bibinfo {year} {2008})}\BibitemShut
  {NoStop}%
\bibitem [{\citenamefont {de~Miranda}\ \emph {et~al.}(2011)\citenamefont
  {de~Miranda}, \citenamefont {Chotia}, \citenamefont {Neyenhuis},
  \citenamefont {Wang}, \citenamefont {QuÃ©mÃ©ner}, \citenamefont {Ospelkaus},
  \citenamefont {Bohn}, \citenamefont {Ye},\ and\ \citenamefont
  {Jin}}]{Miranda2011stereodynamics}%
  \BibitemOpen
  \bibfield  {author} {\bibinfo {author} {\bibfnamefont {M.~H.~G.}\
  \bibnamefont {de~Miranda}}, \bibinfo {author} {\bibfnamefont
  {A.}~\bibnamefont {Chotia}}, \bibinfo {author} {\bibfnamefont
  {B.}~\bibnamefont {Neyenhuis}}, \bibinfo {author} {\bibfnamefont
  {D.}~\bibnamefont {Wang}}, \bibinfo {author} {\bibfnamefont {G.}~\bibnamefont
  {QuÃ©mÃ©ner}}, \bibinfo {author} {\bibfnamefont {S.}~\bibnamefont
  {Ospelkaus}}, \bibinfo {author} {\bibfnamefont {J.~L.}\ \bibnamefont {Bohn}},
  \bibinfo {author} {\bibfnamefont {J.}~\bibnamefont {Ye}}, \ and\ \bibinfo
  {author} {\bibfnamefont {D.~S.}\ \bibnamefont {Jin}},\ }\href
  {https://doi.org/10.1038/nphys1939} {\bibfield  {journal} {\bibinfo
  {journal} {Nature Physics}\ }\textbf {\bibinfo {volume} {7}},\ \bibinfo
  {pages} {502} (\bibinfo {year} {2011})}\BibitemShut {NoStop}%
\bibitem [{\citenamefont {Ye}\ \emph {et~al.}(2018)\citenamefont {Ye},
  \citenamefont {Guo}, \citenamefont {GonzÃ¡lez-MartÃ­nez}, \citenamefont
  {QuÃ©mÃ©ner},\ and\ \citenamefont {Wang}}]{Ye2018NaRb}%
  \BibitemOpen
  \bibfield  {author} {\bibinfo {author} {\bibfnamefont {X.}~\bibnamefont
  {Ye}}, \bibinfo {author} {\bibfnamefont {M.}~\bibnamefont {Guo}}, \bibinfo
  {author} {\bibfnamefont {M.~L.}\ \bibnamefont {GonzÃ¡lez-MartÃ­nez}},
  \bibinfo {author} {\bibfnamefont {G.}~\bibnamefont {QuÃ©mÃ©ner}}, \ and\
  \bibinfo {author} {\bibfnamefont {D.}~\bibnamefont {Wang}},\ }\href
  {http://advances.sciencemag.org/content/4/1/eaaq0083.abstract} {\bibfield
  {journal} {\bibinfo  {journal} {Sci Adv}\ }\textbf {\bibinfo {volume} {4}},\
  \bibinfo {pages} {eaaq0083} (\bibinfo {year} {2018})}\BibitemShut {NoStop}%
\bibitem [{\citenamefont {Gregory}\ \emph {et~al.}(2019)\citenamefont
  {Gregory}, \citenamefont {Frye}, \citenamefont {Blackmore}, \citenamefont
  {Bridge}, \citenamefont {Sawant}, \citenamefont {Hutson},\ and\ \citenamefont
  {Cornish}}]{Gregory2019RbCsSticky}%
  \BibitemOpen
  \bibfield  {author} {\bibinfo {author} {\bibfnamefont {P.~D.}\ \bibnamefont
  {Gregory}}, \bibinfo {author} {\bibfnamefont {M.~D.}\ \bibnamefont {Frye}},
  \bibinfo {author} {\bibfnamefont {J.~A.}\ \bibnamefont {Blackmore}}, \bibinfo
  {author} {\bibfnamefont {E.~M.}\ \bibnamefont {Bridge}}, \bibinfo {author}
  {\bibfnamefont {R.}~\bibnamefont {Sawant}}, \bibinfo {author} {\bibfnamefont
  {J.~M.}\ \bibnamefont {Hutson}}, \ and\ \bibinfo {author} {\bibfnamefont
  {S.~L.}\ \bibnamefont {Cornish}},\ }\href
  {https://doi.org/10.1038/s41467-019-11033-y} {\bibfield  {journal} {\bibinfo
  {journal} {Nature Communications}\ }\textbf {\bibinfo {volume} {10}},\
  \bibinfo {pages} {3104} (\bibinfo {year} {2019})}\BibitemShut {NoStop}%
\bibitem [{\citenamefont {Hu}\ \emph {et~al.}(2019)\citenamefont {Hu},
  \citenamefont {Liu}, \citenamefont {Grimes}, \citenamefont {Lin},
  \citenamefont {Gheorghe}, \citenamefont {Vexiau}, \citenamefont
  {Bouloufa-Maafa}, \citenamefont {Dulieu}, \citenamefont {Rosenband},\ and\
  \citenamefont {Ni}}]{Hu2019KRbReaction}%
  \BibitemOpen
  \bibfield  {author} {\bibinfo {author} {\bibfnamefont {M.-G.}\ \bibnamefont
  {Hu}}, \bibinfo {author} {\bibfnamefont {Y.}~\bibnamefont {Liu}}, \bibinfo
  {author} {\bibfnamefont {D.~D.}\ \bibnamefont {Grimes}}, \bibinfo {author}
  {\bibfnamefont {Y.-W.}\ \bibnamefont {Lin}}, \bibinfo {author} {\bibfnamefont
  {A.~H.}\ \bibnamefont {Gheorghe}}, \bibinfo {author} {\bibfnamefont
  {R.}~\bibnamefont {Vexiau}}, \bibinfo {author} {\bibfnamefont
  {N.}~\bibnamefont {Bouloufa-Maafa}}, \bibinfo {author} {\bibfnamefont
  {O.}~\bibnamefont {Dulieu}}, \bibinfo {author} {\bibfnamefont
  {T.}~\bibnamefont {Rosenband}}, \ and\ \bibinfo {author} {\bibfnamefont
  {K.-K.}\ \bibnamefont {Ni}},\ }\href {\doibase 10.1126/science.aay9531}
  {\bibfield  {journal} {\bibinfo  {journal} {Science}\ }\textbf {\bibinfo
  {volume} {366}},\ \bibinfo {pages} {1111} (\bibinfo {year}
  {2019})}\BibitemShut {NoStop}%
\bibitem [{\citenamefont {Drews}\ \emph {et~al.}(2017)\citenamefont {Drews},
  \citenamefont {DeiÃŸ}, \citenamefont {Jachymski}, \citenamefont {Idziaszek},\
  and\ \citenamefont {Hecker~Denschlag}}]{Drews2017Rb2}%
  \BibitemOpen
  \bibfield  {author} {\bibinfo {author} {\bibfnamefont {B.}~\bibnamefont
  {Drews}}, \bibinfo {author} {\bibfnamefont {M.}~\bibnamefont {DeiÃŸ}},
  \bibinfo {author} {\bibfnamefont {K.}~\bibnamefont {Jachymski}}, \bibinfo
  {author} {\bibfnamefont {Z.}~\bibnamefont {Idziaszek}}, \ and\ \bibinfo
  {author} {\bibfnamefont {J.}~\bibnamefont {Hecker~Denschlag}},\ }\href
  {https://doi.org/10.1038/ncomms14854} {\bibfield  {journal} {\bibinfo
  {journal} {Nature Communications}\ }\textbf {\bibinfo {volume} {8}},\
  \bibinfo {pages} {14854} (\bibinfo {year} {2017})}\BibitemShut {NoStop}%
\bibitem [{\citenamefont {Mayle}\ \emph {et~al.}(2012)\citenamefont {Mayle},
  \citenamefont {Ruzic},\ and\ \citenamefont {Bohn}}]{Mayle2012statscat}%
  \BibitemOpen
  \bibfield  {author} {\bibinfo {author} {\bibfnamefont {M.}~\bibnamefont
  {Mayle}}, \bibinfo {author} {\bibfnamefont {B.~P.}\ \bibnamefont {Ruzic}}, \
  and\ \bibinfo {author} {\bibfnamefont {J.~L.}\ \bibnamefont {Bohn}},\ }\href
  {\doibase 10.1103/PhysRevA.85.062712} {\bibfield  {journal} {\bibinfo
  {journal} {Phys. Rev. A}\ }\textbf {\bibinfo {volume} {85}},\ \bibinfo
  {pages} {062712} (\bibinfo {year} {2012})}\BibitemShut {NoStop}%
\bibitem [{\citenamefont {Mayle}\ \emph {et~al.}(2013)\citenamefont {Mayle},
  \citenamefont {Qu\'em\'ener}, \citenamefont {Ruzic},\ and\ \citenamefont
  {Bohn}}]{Mayle2013resscat}%
  \BibitemOpen
  \bibfield  {author} {\bibinfo {author} {\bibfnamefont {M.}~\bibnamefont
  {Mayle}}, \bibinfo {author} {\bibfnamefont {G.}~\bibnamefont {Qu\'em\'ener}},
  \bibinfo {author} {\bibfnamefont {B.~P.}\ \bibnamefont {Ruzic}}, \ and\
  \bibinfo {author} {\bibfnamefont {J.~L.}\ \bibnamefont {Bohn}},\ }\href
  {\doibase 10.1103/PhysRevA.87.012709} {\bibfield  {journal} {\bibinfo
  {journal} {Phys. Rev. A}\ }\textbf {\bibinfo {volume} {87}},\ \bibinfo
  {pages} {012709} (\bibinfo {year} {2013})}\BibitemShut {NoStop}%
\bibitem [{\citenamefont {Norrgard}\ \emph {et~al.}(2016)\citenamefont
  {Norrgard}, \citenamefont {McCarron}, \citenamefont {Steinecker},
  \citenamefont {Tarbutt},\ and\ \citenamefont {DeMille}}]{norrgard16RF}%
  \BibitemOpen
  \bibfield  {author} {\bibinfo {author} {\bibfnamefont {E.}~\bibnamefont
  {Norrgard}}, \bibinfo {author} {\bibfnamefont {D.}~\bibnamefont {McCarron}},
  \bibinfo {author} {\bibfnamefont {M.}~\bibnamefont {Steinecker}}, \bibinfo
  {author} {\bibfnamefont {M.}~\bibnamefont {Tarbutt}}, \ and\ \bibinfo
  {author} {\bibfnamefont {D.}~\bibnamefont {DeMille}},\ }\href@noop {}
  {\bibfield  {journal} {\bibinfo  {journal} {Phys. Rev. Lett.}\ }\textbf
  {\bibinfo {volume} {116}},\ \bibinfo {pages} {063004} (\bibinfo {year}
  {2016})}\BibitemShut {NoStop}%
\bibitem [{\citenamefont {McCarron}\ \emph {et~al.}(2018)\citenamefont
  {McCarron}, \citenamefont {Steinecker}, \citenamefont {Zhu},\ and\
  \citenamefont {DeMille}}]{McCarron18}%
  \BibitemOpen
  \bibfield  {author} {\bibinfo {author} {\bibfnamefont {D.}~\bibnamefont
  {McCarron}}, \bibinfo {author} {\bibfnamefont {M.}~\bibnamefont
  {Steinecker}}, \bibinfo {author} {\bibfnamefont {Y.}~\bibnamefont {Zhu}}, \
  and\ \bibinfo {author} {\bibfnamefont {D.}~\bibnamefont {DeMille}},\
  }\href@noop {} {\bibfield  {journal} {\bibinfo  {journal} {Physical Review
  Letters}\ }\textbf {\bibinfo {volume} {121}} (\bibinfo {year}
  {2018})}\BibitemShut {NoStop}%
\bibitem [{\citenamefont {Anderegg}\ \emph {et~al.}(2018)\citenamefont
  {Anderegg}, \citenamefont {Augenbraun}, \citenamefont {Bao}, \citenamefont
  {Burchesky}, \citenamefont {Cheuk}, \citenamefont {Ketterle},\ and\
  \citenamefont {Doyle}}]{anderegg18}%
  \BibitemOpen
  \bibfield  {author} {\bibinfo {author} {\bibfnamefont {L.}~\bibnamefont
  {Anderegg}}, \bibinfo {author} {\bibfnamefont {B.~L.}\ \bibnamefont
  {Augenbraun}}, \bibinfo {author} {\bibfnamefont {Y.}~\bibnamefont {Bao}},
  \bibinfo {author} {\bibfnamefont {S.}~\bibnamefont {Burchesky}}, \bibinfo
  {author} {\bibfnamefont {L.~W.}\ \bibnamefont {Cheuk}}, \bibinfo {author}
  {\bibfnamefont {W.}~\bibnamefont {Ketterle}}, \ and\ \bibinfo {author}
  {\bibfnamefont {J.~M.}\ \bibnamefont {Doyle}},\ }\href {\doibase
  10.1038/s41567-018-0191-z} {\bibfield  {journal} {\bibinfo  {journal} {Nature
  Physics}\ }\textbf {\bibinfo {volume} {14}},\ \bibinfo {pages} {890}
  (\bibinfo {year} {2018})}\BibitemShut {NoStop}%
\bibitem [{\citenamefont {Cheuk}\ \emph {et~al.}(2018)\citenamefont {Cheuk},
  \citenamefont {Anderegg}, \citenamefont {Augenbraun}, \citenamefont {Bao},
  \citenamefont {Burchesky}, \citenamefont {Ketterle},\ and\ \citenamefont
  {Doyle}}]{cheuk18}%
  \BibitemOpen
  \bibfield  {author} {\bibinfo {author} {\bibfnamefont {L.~W.}\ \bibnamefont
  {Cheuk}}, \bibinfo {author} {\bibfnamefont {L.}~\bibnamefont {Anderegg}},
  \bibinfo {author} {\bibfnamefont {B.~L.}\ \bibnamefont {Augenbraun}},
  \bibinfo {author} {\bibfnamefont {Y.}~\bibnamefont {Bao}}, \bibinfo {author}
  {\bibfnamefont {S.}~\bibnamefont {Burchesky}}, \bibinfo {author}
  {\bibfnamefont {W.}~\bibnamefont {Ketterle}}, \ and\ \bibinfo {author}
  {\bibfnamefont {J.~M.}\ \bibnamefont {Doyle}},\ }\href
  {http://adsabs.harvard.edu/abs/2018PhRvL.121h3201C} {\bibfield  {journal}
  {\bibinfo  {journal} {Physical Review Letters}\ }\textbf {\bibinfo {volume}
  {121}},\ \bibinfo {pages} {083201} (\bibinfo {year} {2018})}\BibitemShut
  {NoStop}%
\bibitem [{\citenamefont {Anderegg}\ \emph {et~al.}(2019)\citenamefont
  {Anderegg}, \citenamefont {Cheuk}, \citenamefont {Bao}, \citenamefont
  {Burchesky}, \citenamefont {Ketterle}, \citenamefont {Ni},\ and\
  \citenamefont {Doyle}}]{Anderegg2019tweezer}%
  \BibitemOpen
  \bibfield  {author} {\bibinfo {author} {\bibfnamefont {L.}~\bibnamefont
  {Anderegg}}, \bibinfo {author} {\bibfnamefont {L.~W.}\ \bibnamefont {Cheuk}},
  \bibinfo {author} {\bibfnamefont {Y.}~\bibnamefont {Bao}}, \bibinfo {author}
  {\bibfnamefont {S.}~\bibnamefont {Burchesky}}, \bibinfo {author}
  {\bibfnamefont {W.}~\bibnamefont {Ketterle}}, \bibinfo {author}
  {\bibfnamefont {K.-K.}\ \bibnamefont {Ni}}, \ and\ \bibinfo {author}
  {\bibfnamefont {J.~M.}\ \bibnamefont {Doyle}},\ }\href
  {http://science.sciencemag.org/content/365/6458/1156.abstract} {\bibfield
  {journal} {\bibinfo  {journal} {Science}\ }\textbf {\bibinfo {volume}
  {365}},\ \bibinfo {pages} {1156} (\bibinfo {year} {2019})}\BibitemShut
  {NoStop}%
\bibitem [{\citenamefont {Xu}\ \emph {et~al.}(2015)\citenamefont {Xu},
  \citenamefont {Yang}, \citenamefont {Liu}, \citenamefont {He}, \citenamefont
  {Zeng}, \citenamefont {Wang}, \citenamefont {Wang}, \citenamefont {Papoular},
  \citenamefont {Shlyapnikov},\ and\ \citenamefont {Zhan}}]{Xu2015atomtweezer}%
  \BibitemOpen
  \bibfield  {author} {\bibinfo {author} {\bibfnamefont {P.}~\bibnamefont
  {Xu}}, \bibinfo {author} {\bibfnamefont {J.}~\bibnamefont {Yang}}, \bibinfo
  {author} {\bibfnamefont {M.}~\bibnamefont {Liu}}, \bibinfo {author}
  {\bibfnamefont {X.}~\bibnamefont {He}}, \bibinfo {author} {\bibfnamefont
  {Y.}~\bibnamefont {Zeng}}, \bibinfo {author} {\bibfnamefont {K.}~\bibnamefont
  {Wang}}, \bibinfo {author} {\bibfnamefont {J.}~\bibnamefont {Wang}}, \bibinfo
  {author} {\bibfnamefont {D.~J.}\ \bibnamefont {Papoular}}, \bibinfo {author}
  {\bibfnamefont {G.~V.}\ \bibnamefont {Shlyapnikov}}, \ and\ \bibinfo {author}
  {\bibfnamefont {M.}~\bibnamefont {Zhan}},\ }\href
  {https://doi.org/10.1038/ncomms8803} {\bibfield  {journal} {\bibinfo
  {journal} {Nature Communications}\ }\textbf {\bibinfo {volume} {6}},\
  \bibinfo {pages} {7803} (\bibinfo {year} {2015})}\BibitemShut {NoStop}%
\bibitem [{\citenamefont {Liu}\ \emph {et~al.}(2018)\citenamefont {Liu},
  \citenamefont {Hood}, \citenamefont {Yu}, \citenamefont {Zhang},
  \citenamefont {Hutzler}, \citenamefont {Rosenband},\ and\ \citenamefont
  {Ni}}]{Liu2018tweezer}%
  \BibitemOpen
  \bibfield  {author} {\bibinfo {author} {\bibfnamefont {L.~R.}\ \bibnamefont
  {Liu}}, \bibinfo {author} {\bibfnamefont {J.~D.}\ \bibnamefont {Hood}},
  \bibinfo {author} {\bibfnamefont {Y.}~\bibnamefont {Yu}}, \bibinfo {author}
  {\bibfnamefont {J.~T.}\ \bibnamefont {Zhang}}, \bibinfo {author}
  {\bibfnamefont {N.~R.}\ \bibnamefont {Hutzler}}, \bibinfo {author}
  {\bibfnamefont {T.}~\bibnamefont {Rosenband}}, \ and\ \bibinfo {author}
  {\bibfnamefont {K.-K.}\ \bibnamefont {Ni}},\ }\href
  {http://science.sciencemag.org/content/360/6391/900.abstract} {\bibfield
  {journal} {\bibinfo  {journal} {Science}\ }\textbf {\bibinfo {volume}
  {360}},\ \bibinfo {pages} {900} (\bibinfo {year} {2018})}\BibitemShut
  {NoStop}%
\bibitem [{\citenamefont {Sompet}\ \emph {et~al.}(2019)\citenamefont {Sompet},
  \citenamefont {Szigeti}, \citenamefont {Schwartz}, \citenamefont {Bradley},\
  and\ \citenamefont {Andersen}}]{Sompet2019coherent}%
  \BibitemOpen
  \bibfield  {author} {\bibinfo {author} {\bibfnamefont {P.}~\bibnamefont
  {Sompet}}, \bibinfo {author} {\bibfnamefont {S.~S.}\ \bibnamefont {Szigeti}},
  \bibinfo {author} {\bibfnamefont {E.}~\bibnamefont {Schwartz}}, \bibinfo
  {author} {\bibfnamefont {A.~S.}\ \bibnamefont {Bradley}}, \ and\ \bibinfo
  {author} {\bibfnamefont {M.~F.}\ \bibnamefont {Andersen}},\ }\href
  {https://www.ncbi.nlm.nih.gov/pubmed/31015406} {\bibfield  {journal}
  {\bibinfo  {journal} {Nature communications}\ }\textbf {\bibinfo {volume}
  {10}},\ \bibinfo {pages} {1889} (\bibinfo {year} {2019})}\BibitemShut
  {NoStop}%
\bibitem [{\citenamefont {Hood}\ \emph {et~al.}(2019)\citenamefont {Hood},
  \citenamefont {Yu}, \citenamefont {Lin}, \citenamefont {Zhang}, \citenamefont
  {Wang}, \citenamefont {Liu}, \citenamefont {Gao},\ and\ \citenamefont
  {Ni}}]{Hood2019FBspec}%
  \BibitemOpen
  \bibfield  {author} {\bibinfo {author} {\bibfnamefont {J.~D.}\ \bibnamefont
  {Hood}}, \bibinfo {author} {\bibfnamefont {Y.}~\bibnamefont {Yu}}, \bibinfo
  {author} {\bibfnamefont {Y.-W.}\ \bibnamefont {Lin}}, \bibinfo {author}
  {\bibfnamefont {J.~T.}\ \bibnamefont {Zhang}}, \bibinfo {author}
  {\bibfnamefont {K.}~\bibnamefont {Wang}}, \bibinfo {author} {\bibfnamefont
  {L.~R.}\ \bibnamefont {Liu}}, \bibinfo {author} {\bibfnamefont
  {B.}~\bibnamefont {Gao}}, \ and\ \bibinfo {author} {\bibfnamefont {K.-K.}\
  \bibnamefont {Ni}},\ }\href@noop {} {\bibfield  {journal} {\bibinfo
  {journal} {arXiv:1907.11226v1}\ } (\bibinfo {year} {2019})}\BibitemShut
  {NoStop}%
\bibitem [{\citenamefont {Meyer}\ and\ \citenamefont
  {Bohn}(2011{\natexlab{a}})}]{Meyer2011SrF}%
  \BibitemOpen
  \bibfield  {author} {\bibinfo {author} {\bibfnamefont {E.~R.}\ \bibnamefont
  {Meyer}}\ and\ \bibinfo {author} {\bibfnamefont {J.~L.}\ \bibnamefont
  {Bohn}},\ }\href {\doibase 10.1103/PhysRevA.83.032714} {\bibfield  {journal}
  {\bibinfo  {journal} {Phys. Rev. A}\ }\textbf {\bibinfo {volume} {83}},\
  \bibinfo {pages} {032714} (\bibinfo {year} {2011}{\natexlab{a}})}\BibitemShut
  {NoStop}%
\bibitem [{\citenamefont {Karman}()}]{tijsemail}%
  \BibitemOpen
  \bibfield  {author} {\bibinfo {author} {\bibfnamefont {T.}~\bibnamefont
  {Karman}},\ }\href@noop {} {}\bibinfo {howpublished} {private
  communication}\BibitemShut {NoStop}%
\bibitem [{\citenamefont {Idziaszek}\ and\ \citenamefont
  {Julienne}(2010)}]{Idziaszek2010universal}%
  \BibitemOpen
  \bibfield  {author} {\bibinfo {author} {\bibfnamefont {Z.}~\bibnamefont
  {Idziaszek}}\ and\ \bibinfo {author} {\bibfnamefont {P.~S.}\ \bibnamefont
  {Julienne}},\ }\href {\doibase 10.1103/PhysRevLett.104.113202} {\bibfield
  {journal} {\bibinfo  {journal} {Phys. Rev. Lett.}\ }\textbf {\bibinfo
  {volume} {104}},\ \bibinfo {pages} {113202} (\bibinfo {year}
  {2010})}\BibitemShut {NoStop}%
\bibitem [{\citenamefont {Frye}\ \emph {et~al.}(2015)\citenamefont {Frye},
  \citenamefont {Julienne},\ and\ \citenamefont {Hutson}}]{Frye2015universal}%
  \BibitemOpen
  \bibfield  {author} {\bibinfo {author} {\bibfnamefont {M.~D.}\ \bibnamefont
  {Frye}}, \bibinfo {author} {\bibfnamefont {P.~S.}\ \bibnamefont {Julienne}},
  \ and\ \bibinfo {author} {\bibfnamefont {J.~M.}\ \bibnamefont {Hutson}},\
  }\href {\doibase 10.1088/1367-2630/17/4/045019} {\bibfield  {journal}
  {\bibinfo  {journal} {New Journal of Physics}\ }\textbf {\bibinfo {volume}
  {17}},\ \bibinfo {pages} {045019} (\bibinfo {year} {2015})}\BibitemShut
  {NoStop}%
\bibitem [{\citenamefont {Meyer}\ and\ \citenamefont
  {Bohn}(2011{\natexlab{b}})}]{SrFChem}%
  \BibitemOpen
  \bibfield  {author} {\bibinfo {author} {\bibfnamefont {E.~R.}\ \bibnamefont
  {Meyer}}\ and\ \bibinfo {author} {\bibfnamefont {J.~L.}\ \bibnamefont
  {Bohn}},\ }\href {\doibase 10.1103/PhysRevA.83.032714} {\bibfield  {journal}
  {\bibinfo  {journal} {Phys. Rev. A}\ }\textbf {\bibinfo {volume} {83}},\
  \bibinfo {pages} {032714} (\bibinfo {year} {2011}{\natexlab{b}})}\BibitemShut
  {NoStop}%
\bibitem [{\citenamefont {Christianen}\ \emph {et~al.}(2019)\citenamefont
  {Christianen}, \citenamefont {Karman},\ and\ \citenamefont
  {Groenenboom}}]{Christianen2019DOS}%
  \BibitemOpen
  \bibfield  {author} {\bibinfo {author} {\bibfnamefont {A.}~\bibnamefont
  {Christianen}}, \bibinfo {author} {\bibfnamefont {T.}~\bibnamefont {Karman}},
  \ and\ \bibinfo {author} {\bibfnamefont {G.~C.}\ \bibnamefont
  {Groenenboom}},\ }\href {\doibase 10.1103/PhysRevA.100.032708} {\bibfield
  {journal} {\bibinfo  {journal} {Phys. Rev. A}\ }\textbf {\bibinfo {volume}
  {100}},\ \bibinfo {pages} {032708} (\bibinfo {year} {2019})}\BibitemShut
  {NoStop}%
\bibitem [{\citenamefont {Bakr}\ \emph {et~al.}(2011)\citenamefont {Bakr},
  \citenamefont {Preiss}, \citenamefont {Tai}, \citenamefont {Ma},
  \citenamefont {Simon},\ and\ \citenamefont {Greiner}}]{Bakr2011algo}%
  \BibitemOpen
  \bibfield  {author} {\bibinfo {author} {\bibfnamefont {W.~S.}\ \bibnamefont
  {Bakr}}, \bibinfo {author} {\bibfnamefont {P.~M.}\ \bibnamefont {Preiss}},
  \bibinfo {author} {\bibfnamefont {M.~E.}\ \bibnamefont {Tai}}, \bibinfo
  {author} {\bibfnamefont {R.}~\bibnamefont {Ma}}, \bibinfo {author}
  {\bibfnamefont {J.}~\bibnamefont {Simon}}, \ and\ \bibinfo {author}
  {\bibfnamefont {M.}~\bibnamefont {Greiner}},\ }\href
  {https://doi.org/10.1038/nature10668} {\bibfield  {journal} {\bibinfo
  {journal} {Nature}\ }\textbf {\bibinfo {volume} {480}},\ \bibinfo {pages}
  {500} (\bibinfo {year} {2011})}\BibitemShut {NoStop}%
\bibitem [{\citenamefont {Kozyryev}\ \emph {et~al.}(2016)\citenamefont
  {Kozyryev}, \citenamefont {Baum}, \citenamefont {Matsuda},\ and\
  \citenamefont {Doyle}}]{kozyryev16}%
  \BibitemOpen
  \bibfield  {author} {\bibinfo {author} {\bibfnamefont {I.}~\bibnamefont
  {Kozyryev}}, \bibinfo {author} {\bibfnamefont {L.}~\bibnamefont {Baum}},
  \bibinfo {author} {\bibfnamefont {K.}~\bibnamefont {Matsuda}}, \ and\
  \bibinfo {author} {\bibfnamefont {J.~M.}\ \bibnamefont {Doyle}},\ }\href
  {\doibase 10.1002/cphc.201601051} {\bibfield  {journal} {\bibinfo  {journal}
  {{ChemPhysChem}}\ }\textbf {\bibinfo {volume} {17}},\ \bibinfo {pages} {3641}
  (\bibinfo {year} {2016})}\BibitemShut {NoStop}%
\bibitem [{\citenamefont {Blackmore}\ \emph {et~al.}(2018)\citenamefont
  {Blackmore}, \citenamefont {Caldwell}, \citenamefont {Gregory}, \citenamefont
  {Bridge}, \citenamefont {Sawant}, \citenamefont {Aldegunde}, \citenamefont
  {Mur-Petit}, \citenamefont {Jaksch}, \citenamefont {Hutson}, \citenamefont
  {Sauer}, \citenamefont {Tarbutt},\ and\ \citenamefont
  {Cornish}}]{blackmore18}%
  \BibitemOpen
  \bibfield  {author} {\bibinfo {author} {\bibfnamefont {J.~A.}\ \bibnamefont
  {Blackmore}}, \bibinfo {author} {\bibfnamefont {L.}~\bibnamefont {Caldwell}},
  \bibinfo {author} {\bibfnamefont {P.~D.}\ \bibnamefont {Gregory}}, \bibinfo
  {author} {\bibfnamefont {E.~M.}\ \bibnamefont {Bridge}}, \bibinfo {author}
  {\bibfnamefont {R.}~\bibnamefont {Sawant}}, \bibinfo {author} {\bibfnamefont
  {J.}~\bibnamefont {Aldegunde}}, \bibinfo {author} {\bibfnamefont
  {J.}~\bibnamefont {Mur-Petit}}, \bibinfo {author} {\bibfnamefont
  {D.}~\bibnamefont {Jaksch}}, \bibinfo {author} {\bibfnamefont {J.~M.}\
  \bibnamefont {Hutson}}, \bibinfo {author} {\bibfnamefont {B.~E.}\
  \bibnamefont {Sauer}}, \bibinfo {author} {\bibfnamefont {M.~R.}\ \bibnamefont
  {Tarbutt}}, \ and\ \bibinfo {author} {\bibfnamefont {S.~L.}\ \bibnamefont
  {Cornish}},\ }\href@noop {} {\bibfield  {journal} {\bibinfo  {journal}
  {Quantum Science and Technology}\ }\textbf {\bibinfo {volume} {4}},\ \bibinfo
  {pages} {014010} (\bibinfo {year} {2018})}\BibitemShut {NoStop}%
\end{thebibliography}
\end{document}